\documentclass[preprint,aps,prd,superscriptaddress,showpacs]{revtex4}

\usepackage{amsmath}
\usepackage{amssymb}
\usepackage{amsfonts}

\def\C{\mathcal{C}}

\def\K{\mathcal{K}}
\def\H{\mathcal{H}}

\def\kuchar{Kucha\v{r}}

\begin{document}




\title{The conformal factor in the parameter-free construction of spin-gauge variables for gravity}

\author{Charles H.-T. Wang}
\email{c.wang@abdn.ac.uk}
\affiliation{School of Engineering and Physical Sciences,
University of Aberdeen, King's College, Aberdeen AB24 3UE, Scotland}
\affiliation{Space Science and Technology Department,
Rutherford Appleton Laboratory, Didcot, Oxon OX11 0QX, England}


\begin{abstract}
The newly found conformal decomposition in canonical general
relativity is applied to drastically simplify the recently
formulated parameter-free construction of spin-gauge variables for
gravity. The resulting framework preserves many of the main
structures of the existing canonical framework for loop quantum
gravity related to the spin network and Thiemann's regularization.
However, the Barbero-Immirzi parameter is now converted into the
conformal factor as a canonical variable. It behaves like a scalar
field but is somehow non-dynamical since the effective Hamiltonian
constraint does not depend on its momentum. The essential steps of
the mathematical derivation of this parameter-free framework for
the spin-gauge variables of gravity are spelled out. The
implications for the loop quantum gravity programme are
briefly discussed.
\end{abstract}


\maketitle

\section{The role of the conformal factor in canonical gravity}


In recent series of works \cite{Wang2005b, Wang2005c, Wang2005d,
Wang2006}, the canonical theory of general relativity has been
constructed in terms of the conformal equivalence classes of
spatial metrics. The work was motivated by the following two
ideas. First, the conformal 3-geometry, rather than the
3-geometry, may well carry the true dynamics of general relativity
(GR) \cite{York1971, York1972, Wang2005a}. Secondly, the free
parameter responsible for the Barbero-Immirzi ambiguity of the
present loop quantum gravity is of conformal nature, and may be
removed by an extension of the phase space for GR using the
conformal symmetry. This led to the canonical framework in
\cite{Wang2005b, Wang2005c} where the conformal 3-metric, the mean
extrinsic curvature and their respective momenta act as canonical
variables. A new first class constraint, the conformal constraint,
was introduced to offset the conformal redundancy. However, its
quantum implementation appears to be impeded by the complexity of
the formalism where the conformal factor as a key subexpression is
a highly non-polynomial function of other variables. This will
cause the required regularization for quantization hard to
resolve.

Fortunately, it was then found in \cite{Wang2006} that the
conformal factor need not be used this way and can receive the
canonical variable status if a suitable canonical transformation
is performed. A quick way of demonstrating this is as follows:
Starting with the Arnowitt-Deser-Misner (ADM) variables $g_{ab}$
and $p^{ab}$, we introduce the conformally related quantities
$\bar{g}_{ab}$ and $\bar{p}^{ab}$ by
\begin{eqnarray}\label{ggamma1}
g_{ab} &=& \phi^4 \bar{g}_{ab}
\\
p^{ab} &=& \phi^{-4} {\bar{p}}^{ab}
\label{pexpr1}
\end{eqnarray}
using the conformal factor $\phi=\phi(x)$.
Throughout this work, we shall use a bar over a quantity to indicate
that the quantity has been obtained with a rescaling using a power of $\phi$.
The strategy is to turn
$\bar{g}_{ab}$ and $\phi$ into new configuration variables. If the
momentum of $\bar{g}_{ab}$ is $\bar{p}^{ab}$, then the momentum of
$\phi$ must be identified. In addition, there should be an
additional constraint to compensate the conformal redundancy in
\eqref{ggamma1} and \eqref{pexpr1}. To this end, let us calculate
the following:
\begin{eqnarray}
p^{ab}\dot{g}_{ab} &=&
\phi^{-4} {\bar{p}}^{ab}[\phi^4 \dot{\bar{g}}_{ab} + 4\,\phi^3\dot{\phi}\,\bar{g}_{ab}]
\nonumber\\
&=&
{\bar{p}}^{ab}\dot{\bar{g}}_{ab}+{\pi}\,\dot{\phi}
\label{pdotg1}
\end{eqnarray}
where
\begin{equation}\label{pi}
{\pi} := 4\,\phi^{-1}\bar{g}_{ab}{\bar{p}}^{ab} = -8\,\phi^{-1}\mu
K
\end{equation}
using the mean extrinsic curvature $K$
and the scale factor $\mu := \sqrt{\det g_{ab}}$. This implies that the
variables $(\bar{g}_{ab}, {\bar{p}}^{ab}; \phi, {\pi})$ may be
treated as a canonical set if the constraint $\C$ given by
\begin{equation}\label{conformconst1}
\C
:=
{\bar{g}}_{ab}{\bar{p}}^{ab} -\frac14\,\phi\,{\pi}
\end{equation}
vanishes weakly. We have therefore identified
$(\bar{g}_{ab}, {\bar{p}}^{ab}; \phi, {\pi})$ as the new
canonical variables together with
$\C$ as the new (conformal) constraint for GR.
In terms of these variables, the ADM diffeomorphism and
Hamiltonian constraints become
\begin{equation}
\label{Jconstr1}
\H_a
=
-2\, \bar\nabla_b \,\bar{p}^b{}_a
+{\pi}\,\phi_{,a}
+4(\ln\phi)_{,a}\, \C
\end{equation}
\begin{equation}
\label{Hconstr1}
\H
=
\phi^{-6}\bar\mu \,\bar{g}_{a b c d}\, {\bar{p}}^{a b} {\bar{p}}^{c d}
-  \phi^2 \bar\mu \bar{R} + 8\,\bar\mu\, \phi \bar{\Delta} \phi
\end{equation}
respectively. Here we have used
the scale factor $\bar\mu := \sqrt{\det \bar{g}_{ab}}$,
Levi-Civita connection $\bar\nabla$,
Ricci scalar curvature $\bar{R}$ and
Laplacian $\bar\Delta := \bar{g}^{ab}\bar\nabla_a\bar\nabla_b$,
associated with the conformal metric $\bar{g}_{ab}$. In \eqref{Jconstr1}, the last term
$4(\ln\phi)_{,a}\, \C$ can be dropped to define an effective
diffeomorphism constraint \cite{Wang2006}.

\section{Triad variables for gravity}

The canonical
framework in the preceding section will be applied to
obtain a triad formulation of GR with extended conformal symmetry.
Before we carry that out, however, it is useful to briefly review the
standard triad formulation of GR with the canonical variables
$({E}^a_i,{K}^i_a)$. Here ${E}^a_i$ is the
densitized triad and ${K}^i_a$ the extrinsic curvature.
In terms of these variables, the ADM variables take the following forms:
\begin{equation}\label{gabE}
g_{ab} = \mu^2  E^i_a  E^i_b
\end{equation}
\begin{align}\label{p2K}
p^{ab}
&=
\frac{1}{2}\,\mu^{-2}
K^i_c
\left[
E_j^c E_i^{(a} E_j^{b)}
-
E_i^c E_j^{a} E_j^{b}
\right]
\end{align}
where $E^i_a$ is the inverse of ${E}^a_i$.
It follows that
\begin{equation}\label{pg}
p^{ab} {g}_{ab}
=
-{E}^a_i {K}^i_a
\end{equation}
and
\begin{equation}\label{pdotg}
p^{ab} \dot{g}_{ab}
=
-\dot{E}^a_i {K}^i_a
-
\frac{1}{\mu^2}
E_i^a \dot{E}^b_i \K_{a b}
\end{equation}
with the constraint
\begin{equation}\label{gauss0}
\K_{a b}
:=
\mu\, {K}^i_{[a} e^i_{b]}
=
\mu^2 {K}^i_{[a} E^i_{b]}
\end{equation}
to vanish weakly.
This justifies $({K}^i_a, {E}^a_i)$ as canonical variables.
Instead of working with the constraint $\K_{a b}$ it is
more convenient to adopt the equivalent constraint
\begin{equation}\label{Ck}
\C_k
:=
\epsilon^{}_{kij} K^{}_{ai} E^a_{j}
=
-\frac{1}{\mu^2} \epsilon^{}_{kij} \K_{a b} E_i^a E_j^b
\end{equation}
since it generates the rotation of the triad. We shall therefore
refer to $\C_k$ as the ``spin constraint''.

In terms of the canonical variables $({K}^i_a, {E}^a_i)$, the ADM diffeomorphism and
Hamiltonian constraints then become
\begin{equation}
\label{HHa}
\H_a
=
2\,E_k^b \,
\nabla^{}_{[a} K^k_{b]}
-\frac12\,
\epsilon^{}_{ijk}\,E^i_a   E^b_j \,\nabla^{}_b \C^{}_k
\end{equation}
and
\begin{equation}\label{HH}
\H =
-\frac1{2\mu}
K^i_{[a} K^j_{b]} E_{i}^a E_{j}^b
-
\mu R
+
\frac1{8\mu}\,\C_{k}\C_{k}
\end{equation}
respectively.

\section{Conformal treatment of the triad variables}

The rescaling relations
\eqref{ggamma1} and \eqref{pexpr1}
give rise to the conformal triad variables
$(\bar{K}^i_a, \bar{E}_i^a)$
satisfying
\begin{align}
\label{Epsilon_i^a}
E_i^a &= \phi^{4}\, \bar{E}_i^a
\\
\label{Kappa^i_a}
K^i_a
&=
\phi^{-4}\, \bar{K}^i_a
\end{align}
Using \eqref{pi} and and the identities
\begin{equation}\label{Ktr}
K^i_a E_i^a = \bar{K}^i_a \bar{E}_i^a = 2 \mu K
\end{equation}
we can calculate that
\begin{eqnarray}
{E}^a_i \dot{K}^i_a
&=&
\phi^{4}\, \bar{E}_i^a[\phi^{-4} \dot{\bar{K}}^i_a - 4\,\phi^{-5}\dot{\phi}\,{\bar{K}}^i_a]
\nonumber\\
&=&
\bar{E}_i^a \dot{\bar{K}}^i_a
- 4\,\phi^{-1} \bar{E}_i^a {\bar{K}}^i_a \, \dot{\phi}
\nonumber\\
&=&
\bar{E}_i^a \dot{\bar{K}}^i_a
+
{\pi}\,\dot{\phi} .
\label{pdotg2}
\end{eqnarray}
This establishes the variables
$(\bar{K}^i_a, \bar{E}_i^a; \phi, \pi)$
as canonical variables.
By using \eqref{pg}, we see that
the conformal constraint
$\C$ defined in \eqref{conformconst1} now takes the form
\begin{equation}\label{conformconst2}
\C
=
-\bar{K}^i_a \bar{E}_i^a -\frac14\,\phi\,{\pi}
\end{equation}
The spin constraint defined in \eqref{Ck} then becomes
\begin{equation}\label{Ck1}
\C_k
=
\epsilon^{}_{kij} \bar{K}^{}_{a[i} \bar{E}^a_{j]} .
\end{equation}
In terms of the variables
$(\bar{K}^i_a, \bar{E}_i^a; \phi, \pi)$
the conformal metric and its momentum
are given by
\begin{equation}\label{gabE2}
\bar{g}_{ab} = \bar{\mu}^2  \bar{E}^i_a  \bar{E}^i_b
\end{equation}
\begin{align}\label{p2K2}
\bar{p}^{ab}
&=
\bar{\mu}^{-2}
\bar{K}^i_c
\left[
\bar{E}_j^c \bar{E}_i^{(a} \bar{E}_j^{b)}
-
\bar{E}_i^c \bar{E}_j^{a} \bar{E}_j^{b}
\right] .
\end{align}
By substituting \eqref{gabE2}, \eqref{p2K2} into \eqref{Jconstr1}, \eqref{Hconstr1}
and making use of the expressions in \eqref{HHa} and \eqref{HH} with the replacement
$(E_i^a, K^i_a)\rightarrow(\bar{E}_i^a, {\bar{K}}^i_a)$ we get
the diffeomorphism constraint and Hamiltonian constraint respectively in the following forms
\begin{equation}
\label{HHa1}
\H_a
=
2\,\bar{E}_k^b \,
\bar{\nabla}^{}_{[a} \bar{K}^k_{b]}
+{\pi}\,\phi_{,a}
-\frac12\,
\epsilon^{}_{ijk}\,\bar{E}^i_a   \bar{E}^b_j \,\bar{\nabla}^{}_b \C^{}_k
+4(\ln\phi)_{,a}\, \C
\end{equation}
\begin{equation}\label{HH1}
\H =
-\frac1{2}\,\phi^{-6}\bar{\mu}^{-1}
\bar{K}^i_{[a} \bar{K}^j_{b]} \bar{E}_{i}^a \bar{E}_{j}^b
-  \phi^2 \bar{\mu}\, \bar{R} + 8\,\bar{\mu}\, \phi \bar{\Delta} \phi
+
\frac18\,\phi^{-6}\bar{\mu}^{-1}\,\C_{k}\C_{k} .
\end{equation}

\section{Standard spin-gauge formalism}

We now briefly review the
existing real spin-gauge formalism
for gravity \cite{Ashtekar1986, Ashtekar1987, Barbero1995a, Immirzi1997}
before moving on to our final form of the
spin-gauge formalism by assimilating the
the conformal treatment in the preceding section.
For any positive constant $\beta$, introduce the spin
connection
\begin{equation}\label{A^i_a}
\tilde{A}^i_a
:=
\Gamma^i_a + \beta \, K^i_a
=
\Gamma^i_a + \tilde{K}^i_a
\end{equation}
where $\tilde{K}^i_a := \beta K^i_a$. Further, introduce the
scaled triad
\begin{equation}\label{E_i^a}
\tilde{E}_i^a := \beta^{-1}\,E_i^a .
\end{equation}
Here and below, we use the tilde to emphasize a quantity's dependence on
the parameter $\beta$. This parameter is called the Barbero-Immirzi
parameter. It can be show that the
variables $(\tilde{A}^i_a, \tilde{E}_i^a)$ are canonical. They are
used for the existing spin-gauge formalism for gravity.
The curvature of the spin connection $\bar{A}^i_a$ is given by
\begin{equation}\label{Fiab}
\tilde{F}^k_{ab} := 2\, \partial^{}_{[a} \tilde{A}^k_{b]} + \epsilon_{kij} \tilde{A}^i_{a} \tilde{A}^j_{b} .
\end{equation}
Denoting by $\tilde{D}_a$
the covariant derivative associated with $\tilde{A}^i_{a}$,
we can express the spin constraint in the form
of the ``Gauss law'' as:
\begin{equation}\label{Gk}
\tilde\C_k
:=
\tilde{D}_a \tilde{E}^a_k
=
\tilde{E}^a_{k,a} + \epsilon_{kij} \, \tilde{A}_a^i \, \tilde{E}_j^a
=
\nabla_a \tilde{E}^a_k + \epsilon_{kij} \, K_a^i \, E_j^a
=
\C_k .
\end{equation}
In the above the torsion-free condition $\nabla_a \tilde{E}^a_k=0$ for the Levi-Civita spin connection $\nabla_a$
associated with the metric ${g}_{ab}$
has been used.
Using the relations
\begin{equation}\label{FPKEG}
\tilde{F}^k_{ab} \tilde{E}_k^b = 2\,\nabla^{}_{[a} \tilde{K}^k_{b]}E_k^b + \tilde{K}^k_{a}
\tilde\C_k
\end{equation}
and
\begin{equation}\label{epsiFPP}
\left[
\beta^{2}\epsilon_{kij} \tilde{F}^k_{ab}
-2\,\beta^2 \tilde{K}^i_{[a} \tilde{K}^j_{b]}
\right]
\tilde{E}_{i}^a \tilde{E}_{j}^b
=
-\mu^2 R
-
2\,\beta^{2} \tilde{E}_k^c \nabla^{}_c \tilde\C_k
\end{equation}
we see that \eqref{HHa} and \eqref{HH} become
\begin{equation}\label{HHH}
\H_a
=
\tilde{F}^k_{ab} \tilde{E}_k^b
-
\tilde{A}^k_{a} \tilde\C_k
+
\Gamma^k_{a} \tilde\C_k
-\frac12\,
\epsilon^{}_{ijk}\,\tilde{E}^i_a   \tilde{E}^b_j \,\nabla^{}_b \tilde\C_k
\end{equation}
and
\begin{align}\label{HHA}
\H
&=
\beta^{1/2}\tilde{\mu}^{-1} \left[
\epsilon_{i j k}\, \tilde{F}^k_{a b}
-\frac{4\beta^2+1}{2\beta^2}\,\tilde{K}^i_{[a} \tilde{K}^j_{b]} \right] \tilde{E}^a_i \tilde{E}^b_j +
2\,\beta^{1/2}\tilde{\mu}^{-1}\, \tilde{E}_k^c \nabla^{}_c \tilde\C_k +
\frac1{8}\,\beta^{-3/2}\tilde{\mu}^{-1}\tilde\C_{k}\tilde\C_{k} .
\end{align}
Here
the scale factor of the metric defined using the triad $\tilde{E}_i^a$
is denoted by $\tilde\mu = \beta^{-3/2} \mu$. As per \cite{AshtekarLewandowski2004},
we summarize the standard spin-gauge formalism of gravity by listing its
effective Hamiltonian constraint:
\begin{align}\label{HHA1}
\tilde\C_\perp
&=
\beta^{1/2}\tilde{\mu}^{-1} \left[
\epsilon_{i j k}\, \tilde{F}^k_{a b}
-\frac{4\beta^2+1}{2\beta^2}\,\tilde{K}^i_{[a} \tilde{K}^j_{b]} \right] \tilde{E}^a_i \tilde{E}^b_j
\end{align}
and diffeomorphism constraint
\begin{equation}\label{Ca}
\tilde\C_a
=
\tilde{F}^k_{ab} \tilde{E}_k^b
-
\tilde{A}^k_{a} \tilde\C_k
\end{equation}
together with the spin constraint
\begin{equation}\label{Gk1}
\tilde\C_k
:=
\tilde{D}_a \tilde{E}^a_k .
\end{equation}

\section{Parameter-free approach to the spin-gauge formalism using the conformal method}

The parameter dependence of the standard spin-gauge formalism discussed above is due to the fact that
the following inequality:
\begin{equation}\label{}
\H[K^i_a, E_i^a] \neq \H[\beta \, K^i_a, \beta^{-1}E_i^a]
\end{equation}
holds for any constant $\beta\neq1$.
This can be traced back to \kuchar's observation that
the kinetic and potential terms are rescaled differently
under a constant conformal transformation \cite{kuchar1981}.
When the phase space of GR is extended by conformal symmetry, the situation is completely different.
From the expression of $\H$ in \eqref{HH1} it is clear that we do have the following equation:
\begin{equation}\label{}
\H[\bar{K}^i_a, \bar{E}_i^a; \phi, \pi]=\H[\beta\bar{K}^i_a, \beta^{-1}\bar{E}_i^a;\beta^{1/4}\phi,\beta^{-1/4}\pi] .
\end{equation}
Therefore, if we introduce the alternative spin connection variable
\begin{equation}\label{c-Al^i_a}
\bar{A}^i_a :=
\bar{\Gamma}^i_a + \bar{K}^i_a
=
\bar{\Gamma}^i_a + \phi^4 \, K^i_a
\end{equation}
and consider
\begin{equation}\label{c-Pi_i^a}
\bar{E}_i^a = \phi^{-4}\,E_i^a
\end{equation}
as its conjugate momentum, then
unlike the construction of
$\tilde{A}^i_a$ and  $\tilde{E}_i^a$ in  \eqref{A^i_a} and \eqref{E_i^a},
any similar multiplicative constant can be absorbed into the
conformal factor $\phi$ together with an inverse rescaling of $\pi$.
As with the spin gauge treatment in the preceding section,
the curvature of the spin connection $\bar{A}^i_a$ is given by
\begin{equation}\label{c-Fiab}
\bar{F}^k_{ab} := 2\, \partial^{}_{[a} \bar{A}^k_{b]} + \epsilon_{kij} \bar{A}^i_{a} \bar{A}^j_{b} .
\end{equation}
Denoting by $\bar{D}_a$
the covariant derivative associated with $\bar{A}^i_{a}$
we can also express the spin constraint in the form
of the Gauss law as:
\begin{equation}\label{c-Gk}
\bar\C_k
:=
\bar{D}_a \bar{E}^a_k
=
\bar{E}^a_{k,a} + \epsilon_{kij} \, \bar{A}_a^i \, \bar{E}_j^a
=
\bar\nabla_a \bar{E}^a_k + \epsilon_{kij} \, \bar{K}_a^i \, \bar{E}_j^a
=
\C_k .
\end{equation}
Here the torsion-free condition $\bar\nabla_a \bar{E}^a_k=0$ for the Levi-Civita spin connection $\bar\nabla_a$
associated with the conformal metric $\bar{g}_{ab}$ \eqref{gabE2}
has been used.
Using the relations
\begin{equation}\label{c-FPKEG}
\bar{F}^k_{ab} \bar{E}_k^b = 2\,\bar\nabla^{}_{[a} \bar{K}^k_{b]}E_k^b + \bar{K}^k_{a}
\bar\C_k
\end{equation}
and
\begin{equation}\label{c-epsiFPP}
\left[
\epsilon_{kij} \bar{F}^k_{ab}
-2\, \bar{K}^i_{[a} \bar{K}^j_{b]}
\right]
\bar{E}_{i}^a \bar{E}_{j}^b
=
-\mu^2 R
-
2\, \bar{E}_k^c \bar\nabla^{}_c \bar\C_k
\end{equation}
analogous to \eqref{FPKEG} and \eqref{epsiFPP},
we see that \eqref{HHa1} and \eqref{HH1} become
\begin{equation}\label{c-HHa1}
\H_a
=
\bar{F}^k_{ab} \bar{E}_k^b
-
\bar{A}^k_{a} \bar\C_k
+{\pi}\,\phi_{,a}
+
\bar\Gamma^k_{a} \bar\C_k
-\frac12\,
\epsilon^{}_{ijk}\,\bar{E}^i_a   \bar{E}^b_j \,\bar\nabla^{}_b \bar\C_k
\end{equation}
and
\begin{align}\label{c-HHA}
\H
&=
{\phi^2}\bar{\mu}^{-1} \left[
\epsilon_{i j k}\, \bar{F}^k_{a b}
-\frac{4\phi^8+1}{2\phi^8}\,\bar{K}^i_{[a} \bar{K}^j_{b]} \right] \bar{E}^a_i \bar{E}^b_j
+ 8\,\bar{\mu}\, \phi \bar{\Delta} \phi
+
2\,{\phi^2}\bar{\mu}^{-1}\, \bar{E}_k^c \bar\nabla^{}_c \bar\C_k +
\frac1{8}\,\phi^{-6}\bar{\mu}^{-1}\bar\C_{k}\bar\C_{k} .
\end{align}
This leads us to the final expressions for
the
effective Hamiltonian constraint
\begin{align}\label{c-HHA1}
\bar\C_\perp
&=
{\phi^2}\bar{\mu}^{-1} \left[
\epsilon_{i j k}\, \bar{F}^k_{a b}
-\frac{4\phi^8+1}{2\phi^8}\,\bar{K}^i_{[a} \bar{K}^j_{b]} \right] \bar{E}^a_i \bar{E}^b_j
+ 8\,\bar{\mu}\, \phi \bar{\Delta} \phi
\end{align}
effective diffeomorphism constraint
\begin{equation}\label{c-Ca}
\bar\C_a
=
\bar{F}^k_{ab} \bar{E}_k^b
-
\bar{A}^k_{a} \bar\C_k
+{\pi}\,\phi_{,a}
\end{equation}
spin constraint
\begin{equation}\label{c-Gk1}
\bar\C_k
=
\bar{D}_a \bar{E}^a_k
\end{equation}
and conformal constraint
\begin{equation}\label{conformconst3}
\bar\C
=
-\bar{K}^i_a \bar{E}_i^a -\frac14\,\phi\,{\pi} .
\end{equation}

\section{Concluding remarks}

It is readily seen that, apart from ``small'' extra terms
involving $\phi$ and $\pi$, the structures of the first three
constraints in \eqref{c-HHA1}, \eqref{c-Ca} and \eqref{c-Gk1}
above resemble very closely that in \eqref{HHA1}, \eqref{Ca} and
\eqref{Gk1}. In fact, the constraints $\bar\C_\perp, \bar\C_a$ and
$\bar\C_k$ reduce to $\tilde\C_\perp, \tilde\C_a$ and $\tilde\C_k$
on substituting $\phi\rightarrow\beta^{1/4}$. However, it is the
introduction of the canonical variable $\phi$ that makes our
spin-gauge formalism parameter-free. It is interesting to observe
that the effective Hamiltonian $\bar\C_\perp$ is independent of
$\pi$. Consequently, the evolution of $\phi$ is like a gauge
effect and is dictated by the effective diffeomorphism constraint
$\bar\C_a$ and the conformal constraint $\bar\C$. This is
consistent with one of our motivating ideas that the true dynamics
of GR is in the conformal 3-geometry, rather than the conformal
factor. An important technical implication of this is related to
the regularization of the Hamiltonian operator. It is envisaged
that the $(\bar{A}^k_{a},\phi)$-representation is to be used for
quantization. While the spin connection $\bar{A}^k_{a}$ can be treated with
spin-networks, the conformal factor $\phi$ will be treated similar to a coupled scalar
field. The appearance of $\phi$ in the first term in
$\eqref{c-HHA1}$ should not spoil the regularization schemes
developed by Thiemann \cite{Thiemann1996, Thiemann1998}, since
$\phi$ commutes with all operators there just like the
Barbero-Immirzi parameter in the existing loop quantum gravity. It
remains to solve the quantum conformal constraint equation
\begin{equation}\label{}
\bar\C\,\Psi[\bar{A}^k_{a},\phi] = 0 .
\end{equation}
Addressing this problem will require the construction of ``conformally related spin networks''
which will form a subject for future investigation.

\begin{acknowledgments}
I wish to thank
J. F. Barbero,
R. Bingham,
S. Carlip,
A. E. Fischer,
G. Immirzi,
C. J. Isham,
J. T. Mendon\c{c}a,
N. O'Murchadha
and
J. W. York
for stimulating discussions. The work is supported by
the Aberdeen Centre for Applied Dynamics Research and the
CCLRC Centre for Fundamental Physics.
\end{acknowledgments}



\begin{thebibliography}{99}

\bibitem{Wang2005b}
C. H.-T. Wang, ``Conformal geometrodynamics: True degrees of freedom in a truly canonical structure'',
Phys. Rev. D {\bf 71}, 124026 (2005)

\bibitem{Wang2005c}
C. H.-T. Wang,
``Unambiguous spin-gauge formulation of canonical general relativity with conformorphism invariance'',
Phys. Rev. D {\bf 72}, 087501 (2005)

\bibitem{Wang2005d}
C. H.-T. Wang,
``Towards conformal loop quantum gravity'',
Talk given at the 4th Meeting on Constrained Dynamics and Quantum Gravity, Cala Gonone, Sardinia, Italy,
12--16 September 2005,
arXiv:gr-qc/0512023

\bibitem{Wang2006}
C. H.-T. Wang,
``Conformal decomposition in canonical general relativity'', arXiv:gr-qc/0603062

\bibitem{York1971}
J. W. York,
``Gravitational degrees of freedom and the initial-value problem'',
Phys. Rev. Lett. {\bf 26}, 1656 (1971)

\bibitem{York1972}
J. W. York,
``Role of coformal 3-geometry in the dynamics of gravitation'',
Phys. Rev. Lett. {\bf 28}, 1082 (1972)

\bibitem{Wang2005a}
C. H.-T. Wang,
``Nonlinear quantum gravity on the constant mean curvature foliation'',
Class. Quantum Grav. {\bf 22}, 33 (2005)

\bibitem{Ashtekar1986}
A. Ashtekar, ``New variables for classical and quantum gravity'', Phys. Rev. Lett. {\bf 57}, 2244 (1986)
\bibitem{Ashtekar1987}
A. Ashtekar, ``New Hamiltonian formulation of general relativity'', Phys. Rev. D {\bf 36}, 1587 (1987)

\bibitem{Barbero1995a}
J. F. Barbero G.,
``Real Ashtekar variables for Lorentzian signature space-times'',
Phys. Rev. D \textbf{51}, 5507 (1995)
\bibitem{Immirzi1997}
G. Immirzi,
``Real and complex connections for canonical gravity'',
Class. Quantum Grav. \textbf{14}, L177 (1997)

\bibitem{AshtekarLewandowski2004}
A. Ashtekar and J. Lewandowski,
``Background independent quantum gravity: a status report'',
Class. Quantum Grav. {\bf 21}, R53 (2004)

\bibitem{kuchar1981}
K. V. \kuchar{}, ``General relativity: Dynamics without symmetry'', J. Math. Phys. {\bf 22}, 2640 (1981)

\bibitem{Thiemann1996}
T. Thiemann, ``Anomaly-free formulation of non-perturbative, four-dimensional Lorentzian quantum gravity'',
Phys. Lett. B {\bf 380}, 257 (1996)
\bibitem{Thiemann1998}
T. Thiemann, ``Quantum spin dynamics (QSD)'', Class. Quantum Grav. {\bf 15}, 839 (1998)

\end{thebibliography}
\end{document}